\documentclass[sigconf]{acmart}
\AtBeginDocument{%
  \providecommand\BibTeX{{%
    \normalfont B\kern-0.5em{\scshape i\kern-0.25em b}\kern-0.8em\TeX}}}

\usepackage{times}
\usepackage{bbm}
\usepackage{tabularx}
\usepackage{latexsym}
\usepackage{amsmath}

\usepackage[T1]{fontenc}
\usepackage[utf8]{inputenc}
\usepackage{microtype}

\setcopyright{rightsretained}
\acmConference[SIGIR eCom'22]{ACM SIGIR Workshop on eCommerce}{July 15, 2022}{ Madrid, Spain}
\acmYear{2022}
\copyrightyear{2022}
\makeatletter
\renewcommand\@formatdoi[1]{\ignorespaces}
\makeatother
\acmISBN{}

\begin{document}

\title{ListBERT: Learning to Rank E-commerce products with Listwise BERT}

\author{Lakshya Kumar*}
\email{lakshya.kumar@myntra.com}
\affiliation{%
\institution{Myntra Designs Pvt. Ltd.}
\country{India}
}

 \author{Sagnik Sarkar*}
 \email{sagnik.sarkar@myntra.com}
 \affiliation{%
  \institution{Myntra Designs Pvt. Ltd.}
  \country{India}
 }
 
\thanks{*Both authors contributed equally to this research.}

\renewcommand{\shortauthors}{Lakshya and Sagnik, et al.}


\begin{abstract}
Efficient search is a critical component for an e-commerce platform with an innumerable number of products. Every day millions of users search for products pertaining to their needs. Thus, showing the relevant products on the top will enhance the user experience. In this work, we propose a novel approach of fusing a transformer-based model with various listwise loss functions for ranking e-commerce products, given a user query. We pre-train a RoBERTa model over a fashion e-commerce corpus and fine-tune it using different listwise loss functions. Our experiments indicate that the RoBERTa model fine-tuned with an NDCG based surrogate loss function(approxNDCG) achieves an NDCG improvement of \textbf{13.9\%} compared to other popular listwise loss functions like ListNET and ListMLE. The approxNDCG based RoBERTa model also achieves an NDCG improvement of \textbf{20.6\%} compared to the pairwise RankNet based RoBERTa model. We call our methodology of directly optimizing the RoBERTa model in an end-to-end manner with a listwise surrogate loss function as \textbf{ListBERT}. Since there is a low latency requirement in a real-time search setting, we show how these models can be easily adopted by using a knowledge distillation technique to learn a representation-focused student model that can be easily deployed and leads to $\sim$ \textbf{10} times lower ranking latency.
\end{abstract}


\maketitle

\section{Introduction}
In an E-commerce\footnote{In this work, we mean fashion e-commerce but this research work applies to general e-commerce} setting every day millions of users enter different queries to search for their desired products. In order to retain the users over the platform, it becomes essential to serve them relevant product results that are coherent with their query. Also, it becomes important to understand the intent of the query with respect to different products in an e-commerce catalog. There are millions of products in a catalog and serving a relevant set of products requires an understanding of both the query and the product. Recently, the transformer\cite{transformer} based models have gained popularity due to their exemplary performance on various NLP tasks, owing to which they are getting adopted in different industry settings to tackle various problems. One can directly apply the transformer-based models to deeply understand the textual data of the queries and the products and learn good contextual representations. These representations can further help in learning a ranking function to finally serve the products resulting in user satisfaction. In an e-commerce platform, a huge amount of implicit user signals gets captured in the search logs that can be directly leveraged to learn a ranking function. Also in the past work\cite{searchlogdata} it has been shown that learning a ranking function from such implicit signals has more utility. In order to learn a ranking function, different learning-to-rank(LTR) methodologies can be applied but listwise is shown to be the most optimal\cite{approxNDCG}. In this work, we propose a novel approach of ranking e-commerce products given a user query with a transformer-based RoBERTa\cite{Roberta} model. Our model directly learns the interactions between the query and the products within the RoBERTa model and finally optimizes over a listwise surrogate loss function for an effective product ranking. We prepare a dataset from a popular fashion e-commerce platform to fine-tune the RoBERTa model directly with the listwise loss. Our main contributions are as follows: 
\begin{itemize}
    \item Propose a novel approach of end-to-end training of a transformer based model with a listwise surrogate loss function. 
    \item Compare the efficacy of different listwise and pairwise loss functions with respect to the NDCG evaluation metric.
    \item Highlight on the adoption of such models in a low-latency e-commerce search setting using knowledge distillation and training a representation-focused student model.
\end{itemize}

\section{Related Work}
\label{sec:Related Work}
Based on the loss function used, LTR approaches proposed in the literature \cite{ltr_book} can be classified into 3 main categories namely, pointwise \cite{pointwise_1} \cite{pointwise_2}, pairwise \cite{burges2010Overall} and listwise \cite{ListNET} \cite{ListMLE} \cite{approxNDCG}. Empirically, listwise approaches usually perform better than pointwise or pairwise approaches mainly because popular information retrieval metrics like NDCG\cite{ir_metrics} consider entire search results lists at once, unlike pointwise and pairwise approaches. Recently, pre-trained BERT models have been successfully applied to the LTR problem. One of the themes in which BERT models are applied to the ranking problem is the multi-stage ranking setup. \citet{monoBERTduoBert} proposed a multi-stage document ranking setup where the point-wise BERT is applied for filtering the documents that fall below a certain threshold in the first stage. In the second stage, a pairwise BERT model is applied to rank the candidate documents to finally serve the results. This multi-stage document ranking model has shown good performance over standard datasets mainly MS MARCO \cite{MSMARCO} and TREC CAR\cite{TRECCAR}. Various other popular multi-stage ranking architectures are mentioned in \cite{bert_ranking}. Another theme for applying BERT-based models to the ranking problem is to use dense learned representations leveraging efficient approximate nearest neighbor lookup. Models like Sentence-BERT\cite{Sentence-BERT} and DPR\cite{dpr} have been proposed for learning the dense representations for ranking. For e-commerce search use cases, there is a strict latency requirement and in order to reduce the latency several works have been proposed using knowledge distillation.  \citet{TwinBERT} applied knowledge distillation and trained a TwinBERT model in the teacher-student framework for effective and efficient retrieval.\\ In this work, we apply a BERT-based\footnote{BERT-based, transformer-based are used interchangeably.} model and fine-tune it in an end-to-end manner on various listwise loss functions and thereby demonstrate an effective method of ranking products in a fashion e-commerce domain. We also fine-tune the BERT based model using a popular pairwise RankNet\cite{RankNet} loss function in order to compare it with listwise loss functions. We apply a Knowledge Distillation technique \cite{KD} to train a representation focused \cite{representation_focused} BERT model that is more suitable in a low-latency e-commerce search setting.  

 \section{Data Preparation}
\label{sec:Data Preparation}
\subsection{Tokenization and Pre-training Dataset}
\label{pre-training data}
The user click-stream data for the product ranking task is not sufficient to optimize the large number of parameters of BERT based models like RoBERTa. Typically these models are first pre-trained on a large corpus of domain specific text and then fine-tuned for the downstream tasks. Therefore, we first pre-train the RoBERTa model from scratch over a fashion dataset which we call as pre-training dataset. In order to better handle the out-of-vocabulory words and effectively tokenize the model inputs for pre-training as well as fine-tuning, we train the BPE\cite{BPE} tokenizer over the pre-training dataset. Once trained, the BPE tokenizer is used across all the different model variants in order to give tokenized query and product inputs to the model. Our pre-training dataset consists of product descriptions (formed using product attributes) and user queries. The median and p90 of the product description lengths are \textbf{27} and \textbf{19} respectively. We collect user queries entered in a span of \textbf{1} month while filtering out single-word queries or low-frequency queries (entered less than 10 times). The median and p90 of the query lengths are \textbf{3} and \textbf{2} respectively. The overall pre-training dataset consists of over \textbf{5 M} training instances with a mixture of product descriptions and user queries. 

\subsection{Fine-tuning Dataset}
\label{fine-tuning data}
We leverage a massive amount of click-stream data generated from user sessions on our e-commerce platform. We follow the idea presented in \citet{searchlogdata}, to convert the empirical Click-Through-Rate (CTR) of products (appearing in the result list of a query) into graded relevance ratings. The formula for the ground truth relevance ratings using CTR, is as follows:
\begin{equation}
    rel_{ctr}(q,d) = ceil(4.\frac{ctr(q,d)}{max_{d \in D_{q}} ctr(q,d)})
\end{equation}
where $ctr(q,d) = \frac{clicks(q,d)}{impressions(q,d)}$, $D_{q}$ is the set of products to be ranked for a query $q$. We sample \textbf{2} weeks of click-stream data for fine-tuning our models. We consider only those products in the result set of a query that had received at least \textbf{50} impressions to ensure reliable CTR estimates. We limit the result set of a query to the top \textbf{30} products returned by the search engine\footnote{Denotes the system within e-commerce that retrieves the products given a user query.}. The train data consists of $\sim$ \textbf{45k} queries. The median and p90 lengths of the result set of these queries are \textbf{30} and \textbf{17} respectively. The median and p90 of the query lengths are both \textbf{3}. To evaluate the performance of our models we create the test set by sampling \textbf{1} week of click-stream data in a time period which is temporally separated from the train data period by \textbf{2} months in order to ensure there will be less overlap between the train and test queries. \footnote{In actual production settings, the train and test would be typically have a temporal separation of 1 week} The test data consists of $\sim$ \textbf{20k} queries. The median and p90 lengths of the result set of the test set queries are \textbf{30} and \textbf{7} respectively. Since the train and test sets are formed using a temporal split there are queries which are common to both sets. The test set contains $\sim$ \textbf{40} $\%$ new queries compared to the train set. Moreover, the result sets for the common queries were also different in the train and test sets. The median percentage of new products per query to be ranked in the test set for the common queries was $\sim$ \textbf{67} $\%$

\begin{figure}[h]
  \centering
  \includegraphics[scale = 0.26]{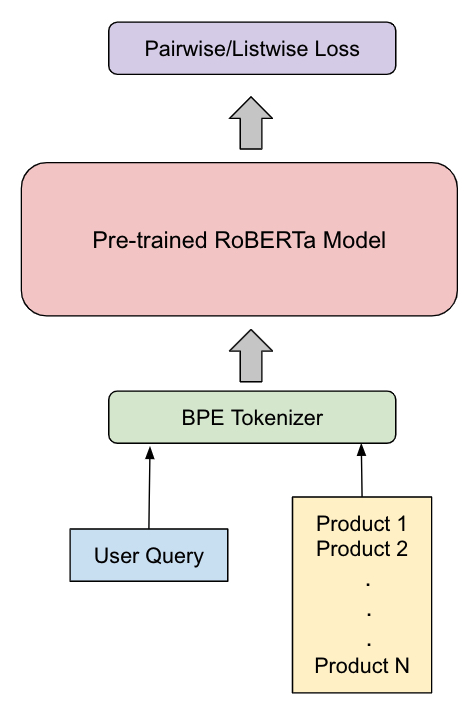}
  \caption{Training setup for ListBERT}
  \label{fig:training_setup}   
\end{figure}

\section{Approach}
\label{sec:approaches}
To learn a good latent representation for both query and product along with the interactions between the query and the product tokens, we use the RoBERTa architecture as our fundamental model. We follow a two-step approach where we first pre-train the RoBERTa model from scratch for adapting to the e-commerce domain and then fine-tune it using pairwise and listwise loss functions for re-ranking of products given a user query. 
\subsection{Pre-training for Domain Knowledge}
The originally proposed pre-trained RoBERTa model\cite{Roberta} has knowledge driven from BookCorpus(16GB)\cite{moviebook}, CC-NEWs(76GB)\cite{NewsDataset}, OpenWebText(38GB)\cite{OpenWebText}, Stories(31GB)\cite{storiesDataset} etc which do not align with the e-commerce queries and products. Therefore, we create a dataset as explained in Section \ref{pre-training data} for pre-training the RoBERTa model in order to incorporate an understanding of the e-commerce domain. We only use the Masked Language Modelling objective\cite{BERT} which is defined as: 
\begin{equation}
    Loss(sentence) = \sum_{i \in MASK}loss(token_{i})
\end{equation}
\begin{equation}
    loss(token_{i}) = -log(p(actualToken_{i}))
    \label{cross-entropyloss}
\end{equation}
where $Loss(sentence)$ corresponds to the overall loss with respect to a sentence in the pre-training dataset. The $loss(token_{i})$ indicates the loss corresponding to each of the MASK tokens. As we randomly mask some of the tokens in a sentence, the loss is computed only with respect to the masked tokens. In equation \ref{cross-entropyloss}, $p(actualToken_{i})$ indicates the probability that the model is assigning to the actual token at position i given the left and the right context tokens. We measure the perplexity after pre-training the model and report the same in Section \ref{sec:experimentsResults}. Finally, we fine-tune this model over a ranking dataset to directly optimize for pairwise and listwise loss functions. 

\subsection{Fine-tuning}
To re-rank the products given a user query, we fine-tune the pre-trained RoBERTa model as shown in Figure \ref{fig:training_setup}.  We optimize a pairwise loss function, i.e., RankNet and three different listwise loss functions, i.e., ListNET\cite{ListNET}, ListMLE\cite{ListMLE} and approxNDCG\cite{approxNDCG}. Finally, we compare these different BERT models on a test dataset as described in \ref{fine-tuning data}.  The formulation for these different loss functions is briefly described in the following subsections. We also describe a knowledge distillation technique where we train a representation focused student model using the ListBERT model as the teacher model in order to reduce the ranking latency.

\subsubsection{\textbf{RankNet}}
RankNet First introduced in \citet{RankNet} using feed forward neural network as the fundamental model. The main objective of RankNet is to perform pairwise ranking optimization by minimizing the number of inversions in the ranked list. An inversion indicates the condition where a low rank product is ranked above a high rank product. In order to solve this optimization problem as described in \citet{allRank}, we first generate the pairs of the products with respect to a given query from the ground truth dataset. Among these pairs, we calculate the true difference between their relevance and retain only those pairs where the difference is positive and finally denote them as $True_{diff}$. From the RoBERTa model, we generate scores for each of the \textbf{N} products\footnote{For every query, we have a ranked list of products which is indicated by N.} with respect to a given query and consider only those pairs which are present in ground truth pair list, i.e., $True_{diff}$ and calculate the difference in their predicted model score and denote them as $Pred_{diff}$. Finally, we apply the below loss function present in \citet{pytorch} for all the queries in the training set to optimize for the pairwise ranking in RankNet.

\begin{equation}
    l(True_{diff},Pred_{diff}) = -\sum_{\substack{i \in True_{diff} \\ j \in Pred_{diff}}}[w_{i} * \mathbbm{1}_{i>0} *log(\sigma(j)))]
\end{equation}
\begin{equation}
    w_i = pow(rel_m, 2) - pow(rel_n, 2)
\end{equation}
where $rel_m$ and $rel_n$ indicates the true relevance of document m and n respectively in a pair i in $True_{diff}$.
\subsubsection{\textbf{ListNET}}
We can assume that we have p user queries which can be denoted as $q^{(i)}$, i = 1,2,....,p and $n_{i}$ denotes the number of products corresponding to the $i$th query. Also, we can assume that $x_{j}^{i}$ corresponds to a feature vector that we obtain using both the query i and the product j. Let $f_{w}$ denote a scoring function that assigns a score given $x_{j}^{i}$. As per \citet{ListNET}, the ListNET loss for a query is given by the following equations: 
\begin{equation}
    P_{z^{i}(f_{w})}(x_{j}^{i}) = \frac{exp(f_{w}(x_{j}^{i}))}{\sum_{k=1}^{n_{i}}exp(f_{w}(x_{k}^{i}))}
    \label{listnet}
\end{equation}
\begin{equation}
    L(y^{i}, z^{i}(f_{w})) = - \sum_{j=1}^{n_i}P_{y^{i}}(x_{j}^{i})log(P_{z^{i}(f_{w})}(x_{j}^{i}))
\end{equation}

Where $z^{i}$ denotes the predicted scores list for a query $q^{(i)}$. And $y^{i}$ denotes the actual scores list as per the true relevance of the products for a given query.

\subsubsection{\textbf{ListMLE}}
\citet{ListMLE} proposed ListMLE loss function, where the likelihood function is given as: 
\begin{equation}
\phi(f_w(x), y) = -log(P(y|x;f_w))
\end{equation}

where $P(\textbf{y}|\textbf{x};f_{w})$ is defined as:
\begin{equation}
   P(y|x;f_{w}) = \prod_{j=1}^{n_{i}}\frac{exp(f_{w}(x_{y(j)}))}{\sum_{k=j}^{n_{i}}exp(f_{w}(x_{y(k)}))}
\end{equation}
where $x_{y(j)}$ indicates the feature vector of a product at position j in the true relevance list. For p queries in the training set, the overall likelihood loss can be given as: 
\begin{equation}
    - \sum_{i=1}^{p}log P(y^{i}|x^{i};f_{w})
\end{equation}
where ${\textbf{y}^{i}}$ corresponds to the actual ranked list of products and \textbf{$\textbf{x}^{i}$} corresponds to the feature vectors for $n^{i}$ products for a given query i.

\subsubsection{\textbf{approxNDCG}}
In order to evaluate a ranking list, NDCG\cite{ir_metrics} is the popular metric that is widely used. But the NDCG cannot be directly optimized owing to its non-differentiability. In order to solve for this, as per \citet{approxNDCG}, position in the DCG\footnote{DCG corresponds to discounted cumulative gain in the NDCG formula.} calculation can be approximated using the below formulation: 
\begin{equation}
    \pi_{f}(i) = 1 + \sum_{j\neq i}\mathbbm{1}_{f(x)|_{i}<f(x)|_{j}}
    \label{positionNDCG}
\end{equation}
where $\mathbbm{1}_{s<t}$ is an indicator function that is 1 when $s<t$ and 0 otherwise. In order to approximate NDCG, the authors proposed the below formulation using the sigmoid function, 
\begin{equation}
    \mathbbm{1}_{s<t} = \mathbbm{1}_{t-s>0}\approx \sigma(t-s) = \frac{1}{1+\exp^{-\alpha(t-s)}}
\end{equation}
In our experiments, RankNet, ListNET, ListMLE, and approxNDCG loss functions are used for fine-tuning the RoBERTa model. After fine-tuning, we apply knowledge distillation to train a representation-focused student model using the best performing $RoBERTa_{NDCG}$ model as a teacher model for low latency ranking. 

\subsubsection{\textbf{Knowledge Distillation}}
\label{sec:KD}
We train a student model consisting of a RoBERTa model with a linear layer at the top. We optimize the student model with a margin Mean Square Error(MSE) loss as proposed in \cite{KD}. The margin MSE loss is given as:
\begin{multline}
  L(Q, P^{+}, P^{-}) = MSE(M_{s}(Q,P^{+}) - M_{s}(Q,P^{-}),\\
                        M_{t}(Q,P^{+}) - M_{t}(Q,P^{-})),
\end{multline}
where Q, P denotes a query and a product respectively. $P^{+}$ and $P^{-}$ corresponds to the high and low relevant products\footnote{$P^{+}$ has higher relevance as compared to $P^{-}$} respectively. $M_{t}$ denotes our teacher model, i.e., $RoBERTa_{NDCG}$ and $M_{s}$ denotes student model. Unlike the teacher model, the student model takes the query and the product text without concatenation. We independently pass the query and the product to obtain their latent representation and then take the dot product to generate a score which is used to optimize the margin MSE loss. As our student model is a representation-focused model, after training it, we can pre-compute and store the representation for all the products in an e-commerce catalog. In the real-time, we generate the query representation on-the-fly from the student model to rank the relevant products by taking the dot product between the query representation and the pre-computed representations of the candidate products. In the next section, we highlight on the performance of the student model and also describe the latency improvement that can be achieved with the representation focused student model. 
\begin{table}[h]
\centering
\begin{tabular}{p{1.95cm}p{1.65cm}p{1.3cm}p{1.3cm}}
\hline
\textbf{Model Type} & \textbf{Surrogate Loss} & \textbf{NDCG} \\
\hline
$RoBERTa_{RNet}$ & RankNet  & 0.625 \\
$RoBERTa_{Net}$ & ListNET  & 0.630 \\
$RoBERTa_{MLE}$ & ListMLE  & 0.662 \\
$RoBERTa_{NDCG}$ & approxNDCG & \textbf{0.754}\\
\hline
\end{tabular}
\caption{\label{results} Ranking Performance of ListBERT models.}
\end{table}
\vspace{0.01cm}
\section{Results \& Experimental Setup}
\label{sec:experimentsResults}
In order to pre-train the RoBERTa model, we first train a Byte-Pair-Encoding(BPE)\cite{BPE} tokenizer over a pre-training dataset(as explained in Section \ref{pre-training data}). The vocabulary size of the tokenizer is \textbf{30K}. Due to the small pre-training data size($\sim$ 6GB) and computation constraints, we use a smaller version of the RoBERTa model having \textbf{6} hidden layers and \textbf{12} attention heads. The input dimension is \textbf{768} and the number of input tokens has a maximum size of \textbf{512}. As we optimize the Masked Language Modelling\cite{BERT} objective, the pre-training data is prepared by randomly masking \textbf{15\%} of the tokens from each of the sentences. For faster pre-training, the multi-GPU pre-training is done with \textbf{2 Tesla V100 GPUs} using Mixed Precision Training\citep{AMP_training}. After pre-training for \textbf{4 epochs}, the perplexity of the RoBERTa model over the test data($\sim$ 1GB)\footnote{This test data is with respect to the pre-training data and consists of user queries and product descriptions.} is \textbf{1.40}. \\
For fine-tuning, we take the pre-trained model and apply a linear layer on top of it. Finally, we experiment with different loss functions as explained in Section \ref{sec:approaches}. We use allRank\cite{allRank} framework-based implementation of pairwise and listwise losses. In order to fine-tune the RoBERTa model, we concatenate both the query and the product description with a \textbf{<SEP>} token and feed it into the RoBERTa model. We take the representation corresponding to the \textbf{<CLS>} token and feed it into the linear layer and optimize for the pairwise and listwise loss functions. For both pre-training and fine-tuning, we use \textbf{Pytorch}\cite{pytorch} framework and \textbf{Huggineface}\cite{wolf2019huggingfaces} based implementation of the RoBERTa model. We use \textbf{Adam}\cite{adam} optimizer to update the parameters in both pre-training and finetuning. The results of different variants of the RoBERTa model after fine-tuning for \textbf{10 epochs} are shown in Table \ref{results}. The RoBERTa model fine-tuned with approxNDCG outperforms RankNet, ListNET and ListMLE loss functions based RoBERTa models with an NDCG of \textbf{0.754}.\\
For knowledge distillation, we train a student RoBERTa model by obtaining the scores from the teacher model\footnote{we take the best teacher model, i.e., $RoBERTa_{NDCG}$} for every query and product pair in the training data\footnote{12M instances.}. Student RoBERTa model takes the query and product independently unlike the teacher which takes the concatenation of query and product. The student model takes the dot product between the query and product representation and then tries to optimize using margin MSE loss as explained in \ref{sec:KD}. The student RoBERTa model achieves an NDCG of \textbf{0.73} over the test data after training for \textbf{4 epochs}. For the teacher model, the average ranking latency is $\sim$ \textbf{154 ms}, whereas for the student model the ranking latency is $\sim$ \textbf{15 ms}. 

\section{Conclusion \& Future Work}
We proposed a novel way of end-to-end training of the transformer-based RoBERTa model in a listwise setting which we call as \textbf{ListBERT}, for ranking e-commerce products. Fine-tuning the RoBERTa model with listwise loss function represents a novel way of learning query and product interactions along with their contextual representations that helps in more optimal ranking of e-commerce products. The RoBERTa model with approxNDCG as the listwise loss outperforms both ListNET and ListMLE with an NDCG of \textbf{0.754}. The ApproxNDCG based RoBERTa model also outperforms the pairwise loss based RoBERTa which we calls as $RoBERTa_{RNet}$. In order to reduce the ranking latency, we trained a representation-focused student model using knowledge distillation that achieves an NDCG of \textbf{0.73} with a 10x latency improvement as compared to the best teacher model, i.e., $RoBERTa_{NDCG}$. In future this work can be extended to incorporate user features for personalized product ranking. It will be interesting to explore different ways to reduce the model size and see the impact of the same on the model performance. Experimenting with other knowledge distillation approaches is another interesting future work.

\bibliographystyle{ACM-Reference-Format}
\bibliography{mainFile}










\end{document}